\documentclass[12pt]{amsart}

\usepackage[dvips]{graphics}

\usepackage{epsfig}

\begin{document}

\title[]{\bf Conformal Anomaly for Free
             Scalar Propagation on Curved Bounded Manifolds}

\author[]{George Tsoupros \\
       {\em The School of Physics,}\\
       {\em Peking University,}\\
       {\em Beijing 100871,}\\
       {\em People's Republic of China}
}
\thanks{present e-mail address: gts@pku.edu.cn}

\begin{abstract}

The trace anomaly for free propagation in the context of a
conformally invariant scalar field theory defined on a curved
manifold of positive constant curvature with boundary is evaluated
through use of an asymptotic heat kernel expansion. In addition to
their direct physical significance the results are also of
relevance to the holographic principle and to Quantum Cosmology.


\end{abstract}

\maketitle


{\bf I. Introduction}\\

The fundamental physical significance of bounded manifolds has
been amply demonstrated in the framework of Euclidean Quantum
Gravity and, more recently, in the context of the holograpic
principle and the $ AdS/CFT$ correspondence. An issue of immediate
importance on such manifolds is the evaluation of the effective
action and, by extension, of the conformal anomaly relevant to the
dynamical behaviour of quantised fields. The contribution to the
conformal anomaly which emerges from free propagation on a curved
manifold is the exclusive result of the gravitational backreaction
on the manifold's geometry and has a distinct character from that
which emerges from matter-to-matter interactions. In the case of
bounded manifolds the conformal anomaly also receives a
simultaneous contribution from the non-trivial boundary. In what
follows a brief outline of the techniques relevant to the
evaluation of the free-propagation-related conformal anomaly on a
general bounded manifold will be presented as the incipient point
of an analysis which advances from the general to the concrete
case of the bounded manifold of positive constant curvature stated
herein.

For reasons of technical convenience the analysis will be
performed on $ C_4$, a segment of the Euclidean sphere bounded by
a hypersurface of positive extrinsic curvature, with homogeneous
Dirichlet-type boundary conditions. Such a choice allows for a
direct use of the results hitherto attained on such a manifold
\cite{G}, \cite{T}, \cite{GT}, \cite{George}. The results obtained
herein have general significance for bounded manifolds of the same
topology both in terms of the general structure of the effective
action and in terms of the interaction between boundary and
surface terms. Notwithstanding that, such results as are obtained
herein on $ C_4$ deserve attention in their own merit due to their
additional significance for the Hartle-Hawking approach to the
quantisation of closed cosmological models.

{\bf II. Trace Anomaly and Free Scalar Propagation on $ C_4$}\\

The scalar component of the bare action defining a theory for a
free, conformal, massless field $ \Phi$ specified on $ C_n$ - a
manifold of positive constant curvature embedded in a $
(n+1)$-dimensional Euclidean space with embedding radius $ a$ and
bounded by a $ (n-1)$-sphere of positive constant extrinsic
curvature $ K$ (diverging normals) - at $ n=4$ is \cite{George}

\begin{equation}
S[\Phi_0] =
\int_{C}d^4\eta\big{[}\frac{1}{2}\Phi_0(\frac{L^2-\frac{1}{2}n(n-2)}{2a^2})\Phi_0
\big{]} + \oint_{\partial C}d^3\eta K\Phi_0^2
\end{equation}
with the subscript $ C$ signifying integration in the interior of
$ C_4$ and with the subscript $ \partial C$ signifying integration
exclusively on its boundary. In either case the subscript $ 4$ has
been omitted as the integration itself renders the associated
dimensionality manifest. In (1) $ \eta$ is the position vector in
the embedding $ (n + 1)$-dimensional Euclidean space signifying
the coordinates $ \eta_{\mu}$ and

$$
L_{\mu \nu} = \eta_{\mu}\frac{\partial}{\partial \eta_{\nu}} -
\eta_{\nu}\frac{\partial}{\partial \eta_{\mu}}
$$
is the generator of rotations. In addition, the Ricci scalar $ R$
relates to the constant embedding radius $ a$ through

\begin{equation}
R = \frac{n(n-1)}{a^2}
\end{equation}
As stated, the bare action (1) is associated by choice with the
homogeneous Dirichlet condition $ \Phi_{|\partial C_4}=0$ for the
scalar field.

The gravitational component of the bare action on $ C_n$ at $n
\rightarrow 4$ is

\begin{equation}
S_g = \frac{1}{16\pi G_0}\int_Cd^4\eta\big{[}2\Lambda_0 - R {]} -
\frac{1}{8\pi G_0}\oint_{\partial C}d^3\eta K
\end{equation}
The boundary term in (3) is essentially the Gibbons-Hawking term
in the gravitational action functional \cite{George}. At one-loop
order it signifies a quantum correction to the Einstein-Hilbert
action which emerges as the result of the influence which the
background curvature and boundary have on free propagation.

In the context of the theory pursued herein the one-loop
vacuum effective action and associated trace anomaly shall be
obtained, in what follows, through use of heat kernel asymptotic
expansions on a general bounded manifold $ M$ \cite{EKP} by
specifying the geometry to be that of $ C_4$ with homogeneous
Dirichlet conditions on $ \partial C_4$ and the coupling between
matter and gravity to be a conformal coupling between a scalar
field and the stated geometry.

The one-loop effective action $ W_0$ associated with the free
scalar propagation on a general manifold $ M$ is, generally, given
by the expression

\begin{equation}
W_0 = \frac{1}{2}TrlnD
\end{equation}
where $ D$ is the operator associated with the scalar propagator
on $ M$, acting on an abstract Hilbert space of states $ |n>$
subject to orthonormality conditions with eigenvalues $ \lambda_n$
\cite{Birrel}. Introducing the generalised $ \zeta$-function as
\cite{EKP}

\begin{equation}
\zeta(s) \equiv Tr[D^{-s}] = \sum_{n}\lambda_n^{-s}
\end{equation}
it follows that in the case of the scalar field it is

\begin{equation}
W_0 = -\frac{1}{2}\frac{\partial}{\partial s}\zeta(s)_{|s=0} -
\frac{1}{2}\zeta(0)ln(\mu^2)
\end{equation}
On the grounds of general theoretical considerations the mean
value of the stress energy-momentum tensor in some vacuum state is

\begin{equation}
<T_{\mu\nu}> = \frac{2}{\sqrt{-g}}\frac{\delta W_0}{\delta
g^{\mu\nu}}
\end{equation}

Such counterterms contained perturbatively in the bare
gravitational action as are necessary to cancel the divergences
which appear in $ W_0$ on a general manifold $ M$ are local in the
metric field and conformally invariant in four dimensions. In
effect, the trace of the renormalised stress tensor in (7)
receives a contribution from $ W_0$ which can be seen from (6) to
relate to $ \zeta(s)$ through \cite{EKP}

\begin{equation}
\int_Md^4x\sqrt{-g}<T^c_c>_r = \zeta(0)
\end{equation}
On a general manifold $ M$, for that matter, the conformal anomaly
emerging from free propagation at one-loop level for conformally
invariant theories is specified by $ \zeta(0)$. This result
remains valid in the presence of a non-trivial boundary on the
understanding that integration over $ M$ also includes $ \partial
M$.

In order to evaluate the trace anomaly associated with free
propagation on $ M$ it is necessary to consider the asymptotic
expansion \cite{EKP}

\begin{equation}
G(t) \sim \sum_{k=0}^{\infty}A_{k}t^{\frac{k - 4}{2}}; ~~~t
\rightarrow 0^{+}
\end{equation}
of the supertrace

\begin{equation}
G(t) = \int_CtrK_D(x, x,t)d^4x = \sum_ne^{-\lambda_nt} = Tre^{-tD}
\end{equation}
of the heat kernel

$$ K_D(x, x',t) = \sum_n<n|x'><x|n>e^{-\lambda_nt} $$
associated with the bounded elliptic operator $ D$ in (4) through
the heat equation

\begin{equation}
(\frac{\partial}{\partial t} + D)K_D(x, x',t) = 0
\end{equation}
The supertrace relates to $ \zeta(s)$ through an inverse Mellin
transform as

\begin{equation}
\zeta(s) = \frac{1}{\Gamma(s)}\int_0^{\infty}t^{s-1}G(t)dt
\end{equation}
The asymptotic expansion in (9) yields, in the context of (12),
the result

\begin{equation}
\zeta(0) = A_4
\end{equation}
which, as (8) reveals, reduces the issue of the conformal anomaly
due to free propagation of matter on $ M$ to the issue of
evaluating the constant coefficient $ A_4$ in (9).

The general asymptotic expansion in (9) is characterised by the
exclusive presence of even-order coefficients $ A_{2k}$ on any
manifold $ M$ for which $ \partial M=0$. The presence of a
non-trivial $ \partial M$ has the effect of generating an
additional boundary-related component for each even-order
coefficient as well as non-vanishing boundary-related values for
all $ A_{2k+1}$ in (9). In general, the coefficients for the
supertrace of the heat kernel associated with the relevant
elliptic operator on a bounded four-dimensional manifold $ M$
admit, in the context of (9), the form

\begin{equation}
A_{2k} = \int_Ma^{(0)}_{2k}\sqrt{g}d^4x + \int_{\partial
M}a^{(1)}_{2k}\sqrt{h}d^3x
\end{equation}

\begin{equation}
A_{2k+1} = \int_{\partial M}a^{(1)}_{2k+1}\sqrt{h}d^3x
\end{equation}
with $ h$ being the induced metric on the boundary.

The local interior coefficients $ a^{(0)}_{2k}$ are specified by
the same local invariants as in the unbounded manifold of the same
local geometry and do not, for that matter, depend on the boundary
conditions. The far more complicated boundary coefficients $
a^{(1)}_k$ necessitate knowledge of the geometry of $ \partial M$
and of the boundary conditions on it in addition to knowledge of
the geometry of $ M$.

If $ M$ is specified to be a Riemannian manifold of positive
constant curvature then $ M$ reduces to $ S_4$ in the absence of a
boundary and to $ C_4$ if $ \partial C_4$ is, itself, specified to
be a Euclidean three-sphere of constant extrinsic cuvature. In
either case, the elliptic operator $ D$ in (4) associated with
free propagation of a massless scalar field conformally coupled to
the background metric is the operator which appears as the
d'Alembertian in (1)

\begin{equation}
D = \frac{L^2-\frac{1}{2}n(n-2)}{2a^2}
\end{equation}
This elliptic operator is unbounded on the Euclidean de Sitter
space $ S_4$ and its zeta-function evaluation results in the
one-loop effective action \cite{Collins}, \cite{DrummondShore}

\begin{equation}
W_0 = \frac{1}{90}\frac{1}{\epsilon} + O(\epsilon^0);~~~ \epsilon
= 4 - n
\end{equation}
for a conformal scalar field with an associated anomalous trace
contribution

\begin{equation}
<T^c_c>_r = -\frac{1}{90}\frac{1}{a^4\Omega_5}
\end{equation}

The elliptic operator in (17) is bounded on $ C_4$. The evaluation
of the trace anomaly due to free scalar propagation on that
manifold necessitates the asymptotic expansion (9) of the
supertrace

\begin{equation}
G(t) = \int_CtrK_D(\eta,\eta,t)d^4\eta
\end{equation}
of the heat kernel $ K_D(\eta,\eta',t)$ associated with the
bounded elliptic operator $ D$ in (16).

It is worth emphasising that, despite appearances stemming from
the homogeneous Dirichlet condition $ \Phi_{|\partial C_4}=0$, the
boundary term

$$
\oint_{\partial C}d^3\eta K\Phi_0^2 $$ in (1) does not vanish.
Such a non-vanishing effect arises as a result of the boundary
condition

\begin{equation}
K_D(\eta,\eta',t=0) = \delta^{(4)}(\eta-\eta')
\end{equation}
- imposed on the solution to the heat equation on $ C_4$

\begin{equation}
(\frac{\partial}{\partial t} + D)K_D(\eta,\eta',t) = 0
\end{equation}
- which offsets the effect of the homogeneous Dirichlet condition
on $ \partial C_4$ \cite{BarvSol}.

The expressions (14) and (15) for the expansion coefficients $
A_k$ in (9) reduce, respectively, to

\begin{equation}
A_{2k}(D,C_4) = \int_Cc^{(0)}_{2k}d^4\eta + \int_{\partial
C}c^{(1)}_{2k}d^3\eta_{B}
\end{equation}
and

\begin{equation}
A_{2k+1}(D,C_4) = \int_{\partial C}c^{(1)}_{2k+1}d^3\eta_B
\end{equation}
on $C_4$, with $ D$ specified in (16) and with the embedding
coordinate vector $ \eta_{B}$ specifying the spherical boundary
hypersurface of maximum colatitude $ \theta_0$. As stated in the
context of (8) and (13), the trace anomaly on $ C_4$ is associated
with the $ A_4(D,C_4)$ coefficient in (22).

If the curvature of a Riemannian manifold $ M$ satisfies the
vacuum Einstein equations with a cosmological constant

\begin{equation}
R_{\mu\nu} = \Lambda g_{\mu\nu}
\end{equation}
then the coefficients $ a^{(0)}_{4}$ and $ a^{(1)}_{4}$ specifying
$ A_4$ in (14) will admit the expressions \cite{MossPol}

\begin{equation}
a^{(0)}_{4} = \alpha_0\Lambda^2 + \alpha_2R_{abcd}R^{abcd}
\end{equation}
and

\begin{equation}
a^{(1)}_{4} = \beta_1\Lambda k + \beta_2 k^3 + \beta_3
kk_{ab}k^{ab} + \beta_4k_a^bk_b^ck_c^a +
\beta_5C_{abcd}k^{ac}n^bn^d
\end{equation}
respectively, where in (26) $ k_{ab}$ is the extrinsic curvature
of $ \partial M$ and $ n$ is the vector normal to $ \partial M$.
The expressions in (25) and (26) essentially disentangle the
geometry-related contributions to the $ A_4$ coefficient from
those contributions which depend on the elliptic operator and
boundary conditions. Specifically, the coefficients $ \alpha_0$
and $ \alpha_2$ multiplying respectively the geometry-dependent
expressions in (25) depend exclusively on the operator in whose
heat-kernel asymptotic expansion $ A_4$ is the constant
coefficient. Likewise, the five coefficients $ \beta_i$ which
respectively multiply the geometry-related expressions in (26)
depend only on the same operator and the conditions specified on
the boundary.

If, in the case of $ \Lambda > 0$, a boundary condition imposed on
(24) is that of a compact four-geometry then the solution to (24)
can be either the spherical cap $ C_4$ or the Euclidean
four-sphere $ S_4$. The former case emerges if the remaining
boundary condition corresponds to the specification of the induced
three-geometry as a three-sphere. The latter case emerges if the
remaining boundary condition corresponds to the absence of a
boundary. In addition, the former case reduces to a disk $
\mathcal{D}$ at the limit of boundary three-spheres small enough
to allow for their embedding in flat Euclidean four-space. These
three solutions are aspects of the Hartle-Hawking no-boundary
proposal for the quantisation of closed universes \cite{HarHawk}.
For the stated boundary conditions these solutions to (24) also
coincide with the corresponding solutions to the Euclidean
Einstein field equations in the presence of a massless scalar
field conformally coupled to gravity on the additional Dirichlet
condition of a constant field on $ \partial C_4$, in the present
case of $ C_4$ as well as in that of $ \mathcal{D}$. Such a
coincidence is a consequence of a vanishing stress tensor for the
conformal scalar field \cite{HarHawk}.

In effect, the constant coefficient $ A_4(D,C_4)$ in the heat
kernel asymptotic expansion for a conformal scalar field on $
C_4$, the corresponding $ A_4(D,S)$ on $ S_4$, as well as the
corresponding coefficient $ A_4(D,\mathcal{D})$ on $ \mathcal{D}$
are expected to be inherently related. With $ \theta_0$ being the
maximum colatitude on $ C_4$, which for that matter specifies $
\partial C_4$, the stated relation is \cite{MossPol}

\begin{equation}
A_4(D,C_4) = A_4(D,S)(\frac{1}{2} - \frac{3}{4}cos\theta_0 +
\frac{1}{4}cos^3\theta_0) + A_4(D,\mathcal{D})cos^3\theta_0 + \
\frac{9}{8}\beta_1cos\theta_0sin^2\theta_0
\end{equation}
where, in conformity with (8), (13) and (18), it is

\begin{equation}
A_4(D,S) = -\frac{1}{90}
\end{equation}
Moreover, the corresponding value in the case of $
\Phi_{|\partial\mathcal{D}}=0$ is \cite{EKP}

\begin{equation}
A_4(D,\mathcal{D}) = - \frac{1}{180}
\end{equation}
and the value of the coefficient $ \beta_1$ for the present case
of $ C_4$ with $ \Phi_{|\partial C_4}=0$ is \cite{MossPol}

\begin{equation}
\beta_1 = \frac{29}{135}
\end{equation}
In effect, equation (27) yields $ A_4(D,C_4)$ for free scalar
propagation.

In the context of (8) and (13) the result which (27)-(30) signify
relates to the conformal anomaly through

\begin{equation}
\int_C d^4\eta <T^c_c>_r^{(C)} + \int_{\partial C} d^3\eta
<T^c_c>_r^{(\partial C)} = A_4(D, C_4)
\end{equation}

In order to arrive at a local expression for the trace anomaly on
$ C_4$ use will be made of the stated fact that on any bounded
manifold the local interior coefficients $ a_{2k}^{(0)}$,
associated through (14) with the asymptotic expansion of the
supertrace of the heat kernel in (9), are specified by the same
local invariants as in the unbounded manifold of the same local
geometry. This, in turn, reveals in the context of (9) that the
local asymptotic expansion of the heat kernel associated with the
operator $ D$ in (16) exclusively in the interior of $ C_4$ is in
coincidence with the local asymptotic expansion of the heat kernel
for the same operator on $ S_4$ so that (22) yields

\begin{equation}
c_{2k}^{(0)} = s_{2k}^{(0)}
\end{equation}
with $ s_{2k}^{(0)}$ being the local coefficients $ a_{2k}^{(0)}$
if $ M$ in (14) is specified as $ S_4$. Setting $ k=2$ and
integrating in the interior of $ C_4$ yields

\begin{equation}
\int_Cd^4\eta c_{4}^{(0)} = \int_Cd^4\eta s_{4}^{(0)}
\end{equation}
and, through (31) and (22)

\begin{equation}
<T_c^c>_r^{(C)}  = s_{4}^{(0)}
\end{equation}
In the context of (14), however, (18) amounts to

\begin{equation}
\int_Sd^4\eta <T_c^c>_r  = -\frac{1}{90} = \int_Sd^4\eta
s_{4}^{(0)}
\end{equation}
so that in view of the constancy of $ s_{2k}^{(0)}$ on $ S_4$ it
is

\begin{equation}
s_{4}^{(0)} = -\frac{1}{90}\frac{1}{a^4\Omega_5}
\end{equation}
Substituting this result in (34) yields

\begin{equation}
<T_c^c>_r^{(C)}  = -\frac{1}{90}\frac{1}{a^4\Omega_5}
\end{equation}
This is the desired local expression for the trace anomaly in the
interior of $ C_4$. As expected, it coincides with the
corresponding expression in (18) for the trace anomaly on $ S_4$.

Finally, substituting (37) in (31) yields

\begin{equation}
\int_{\partial C} d^3\eta <T^c_c>_r^{(\partial C)} = A_4(D, C_4) +
\frac{1}{90}(-\frac{1}{3}sin^2\theta_4^0cos\theta_4^0 -
\frac{2}{3}cos\theta_4^0 + \frac{2}{3})
\end{equation}
with \cite{G}

\begin{equation}
\int_Cd^4\eta = a^42\pi^2(-\frac{1}{3}sin^2\theta_4^0cos\theta_4^0
- \frac{2}{3}cos\theta_4^0 + \frac{2}{3})
\end{equation}
Again, the boundary-related contribution $ <T^c_c>_r^{(\partial
C)}$ is constant on $ \partial C_4 \equiv S_3$. Consequently,

\begin{equation}
<T_c^c>_r^{(\partial C)} = \frac{1}{a^34\pi}A_4(D, C_4) +
\frac{1}{90}\frac{1}{a^34\pi}(-\frac{1}{3}sin^2\theta_4^0cos\theta_4^0
- \frac{2}{3}cos\theta_4^0 + \frac{2}{3})
\end{equation}
This is the desired local expression for the trace anomaly on $
\partial C_4$. As expected, it is contingent upon the specified
homogeneous Dirichlet condition through (27)-(30).

In addition to their direct physical significance for the
dynamical behaviour of scalar fields on bounded manifolds such
results as have been obtained herein are also of relevance to the
holographic principle and to Quantum Cosmology.

\end{document}